\documentclass[aip,twocolumn,10pt,superscriptaddress,amsfonts,amssymb,amsmath]{revtex4-1}
\usepackage[active]{srcltx}
\usepackage{epsf}
\usepackage{bm}
\usepackage{multirow}
\usepackage{graphicx}
\usepackage{dcolumn}
\usepackage{color}

\newcommand{\be}{\begin{equation}}
\newcommand{\ee}{\end{equation}}
\newcommand{\bfig}{\begin{figure}}
\newcommand{\efig}{\end{figure}}

\newcommand{\g}{graphene}

\usepackage{bm}
\begin{document}

\title{Functionalization of edge reconstructed graphene nanoribbons by H and Fe: a density functional study}

\author{Soumyajyoti Haldar}
\affiliation{Department of Physics and Astronomy, Uppsala University, Box 516, 75120 Uppsala, Sweden}

\author{Sumanta Bhandary}
\affiliation{Department of Physics and Astronomy, Uppsala University, Box 516, 75120 Uppsala, Sweden}

\author{Satadeep Bhattacharjee}
\affiliation{Department of Physics and Astronomy, Uppsala University, Box 516, 75120 Uppsala, Sweden}

\author{Olle Eriksson}
\affiliation{Department of Physics and Astronomy, Uppsala University, Box 516, 75120 Uppsala, Sweden}

\author{Dilip Kanhere}
\affiliation{Department of Physics, Central University of Rajasthan, Bander Sindri Campus, Dist-Ajmer, Rajasthan-305801, India}

\author{Biplab Sanyal}
\affiliation{Department of Physics and Astronomy, Uppsala University, Box 516, 75120 Uppsala, Sweden}

%\date{\today}

\begin{abstract} 
In this paper, we have studied functionalization of 5-7 edge-reconstructed graphene nanoribbons by ab initio density functional calculations. Our studies show that hydrogenation at the reconstructed edges is favorable in contrast to the case of unreconstructed  6-6 zigzag edges, in agreement with previous theoretical results. Thermodynamical calculations reveal the relative stability of single and dihydrogenated edges under different temperatures and chemical potential of hydrogen gas. From phonon calculations, we find that the lowest optical phonon modes are hardened due to 5-7 edge reconstruction compared to the 6-6 unreconstructed hydrogenated edges. Finally, edge functionalization by Fe atoms reveals a dimerized Fe chain structure along the edges. The magnetic exchange coupling across the edges varies between ferromagnetic and antiferromagnetic ones with the variation of the width of the nanoribbons.
\end{abstract}

\maketitle

\section{Introduction} 

Graphene is a wonder material with many extraordinary electronic, mechanical and optical
properties to be used in future technology \cite{graphene1, graphene3}.
Moreover, the similarity between its properties and the phenomena observed in high energy
physics has created enormous possibilities to test fundamental theories in quantum
electrodynamics by table-top experiments. Apart from exploring the beautiful physics
associated with the Dirac cones in the Brillouin zone, a perpetual interest exists in the
chemical functionalization of graphene to realize new properties. One of the routes of
chemical functionalization is through the creation of defects in graphene and hence,
modification of its properties \cite{chemfunc1,chemfunc2,chemfunc3,bhandaryprl,erik}. The other notable
effort is to attach chemical species (H, F etc.)  to graphene to open up band gaps by
altering sp$^2$ bonds to sp$^3$ ones between C atoms \cite{grapxene1,grapxene2,grapxene3}.
Very recently, it has also been shown \cite{bnc1,bnc2,bnc3} that a combination of boron nitride and
carbon in a two dimensional network can yield interesting electronic properties, e.g.,
opening of band gaps in an otherwise zero band gap semiconducting situation in pure
graphene.

Graphene nanoribbons (GNRs) have attracted a lot of attention in the last few years as
they are potential candidates for future nanoelectronics. It is well-known \cite{son} that
armchair nanoribbons are semiconductors while the zigzag GNRs (ZGNRs) have magnetic edges
coupled to each other antiferromagnetically to open up a gap. Band gap engineering as a
function of the thickness of GNRs is an important study towards realizing tunable
electronics \cite{philipkimprl}. Also, chemical functionalization of GNR edges to achieve
novel properties is another strong motivation to study GNRs. Recent theoretical studies
have reported the possibility of realizing zigzag and armchair type nanoribbons at the interface of
graphene and graphane (hydrogenated graphene) \cite{aksingh, prachi, haldar,zhang} demonstrating an interesting way of generating nanoribbons.

It has been proposed that apart from realizing the conventional and most abundant GNR
edges, viz., zigzag and armchair, one may consider self reconstructed edges
\cite{koskinenprl} where the hexagonal rings at the edges reconstruct to form
pentagon-heptagonal pairs (reczag). This is similar to what has been observed as 5-7-5 Stone-Wales
defects in bulk graphene. Calculations suggest that the total energy of a reconstructed edge is 0.35
eV/\AA~lower than that of a zigzag one. The presence of a reconstructed edge geometry has
been confirmed in experiments \cite{girit} using aberration corrected high resolution
transmission electron microscopy. A recent review article \cite{revedges} has discussed
about the formation of defective edges along with the folded ones where the adjacent
edges of multilayered graphene can join to form closed loops. The discussions on the
zigzag and reconstructed edges is very important from the point of view of magnetism at
the GNR edges, which is a debatable issue.  Recent density functional calculations
\cite{pati} suggest that the single edge reconstructed GNRs show magnetism with
metallic edges although the reconstructions allowed at both edges do not show any
magnetism. Rodrigues {\it et al.} \cite{rodrigues} have studied reconstructed zigzag nanoribbons decorated by Stone-Wales defects by tight-binding theory with the parameters extracted from first principles electronic structure calculations.

From the above discussions, it is clear that the formation of reconstructed edges can
modify the properties of GNRs drastically. So, a thorough understanding of the properties
of the edges is essential along with the possibilities of realizing novel properties due to chemical
functionalization by adatoms or molecules \cite{funcmol}.  %Though a vast literature exists on the functionalization, especially,
%hydrogenation at the edges of zigzag and armchair nanoribbons,  a systematic study of the
%effect of chemical and magnetic functionalization of reczag edges is missing. 
The motivation of this present work is to study of the properties of edge-functionalized reconstructed GNRs by ab initio calculations.
We have focussed on the geometries and electronic structures of reczag edges of varying thicknesses, their stability at finite temperatures under hydrogenation. Finally, we have exploited the possibility of realizing magnetism by decorating the edges with Fe atoms. In this regard, electronic structure, magnetic exchange coupling across the edges and the stable geometries of Fe chains at the edges have been studied.

\section{Computational Details}

First principles spin-polarized density functional calculations were performed using a
plane-wave projector augmented wave (PAW) method based code, VASP~\cite{vasp1,vasp2}.  The
generalized gradient approximation (GGA) as proposed by Perdew, Burke and Ernzerhof 
(PBE) \cite{pbe} was used for the exchange-correlation functional. We have considered different sizes
of double edged reczag, which were infinite along the y axis. To create a sufficient vacuum
in order to avoid the interactions within adjacent cells, unit cell dimensions along x and
z axis were considered as 50 \AA~ and 16 \AA~ respectively. The electronic wave functions
were expanded using plane waves up to a kinetic energy of 500 eV. The electron smearing
used was Fermi smearing with a broadening of 0.05 eV. The energy and the Hellman-Feynman
force thresholds were kept at 10$^{-5}$ eV and 0.005 eV/{\AA} respectively. All atomic
positions were allowed to relax and the elemental cell was kept at a constant size during
the optimization. For all electronic and ionic calculations, we used a 1 $\times$ 60 $\times$ 1 {\bf $k$} Monkhorst-Pack k-point mesh, whereas for phonon calculations, a $1\times 79\times 1$ mesh was used. The phonon frequencies and displacements were obtained using frozen phonon
method \cite{Martin}. A few selected calculations for the geometry optimizations and electronic structures were repeated by using the Quantum Espresso code \cite{qe} using plane wave basis sets and pseudopotentials within GGA-PBE. We tested the
convergence with respect to the basis-set cutoff energy and a value of 80 Ry. was
considered for all results shown in this paper. The other parameters were similar to those
used in VASP calculations. We found out that the results obtained by the two codes are quite
similar. 

In all our calculations, we have kept the unit cell vectors fixed. However, we have tested cell relaxation for 4-rows 2H terminated reczag GNR, as in this case, the effect of cell relaxation is expected to be the largest. From our calculations, we find that the change in unit cell length along the periodic direction (y) is less than 3 \% and along x direction, it is 0.7 \%. The maximum change in the C-C bond length is 1.8 \% whereas the maximum change in the H-C-H bond angle is only 0.2 \%. Our results are not affected by these changes due to cell relaxation. 

\section{Results}

\subsection{Edge termination by H}

\begin{figure*}
\begin{center}
\includegraphics[scale=0.32]{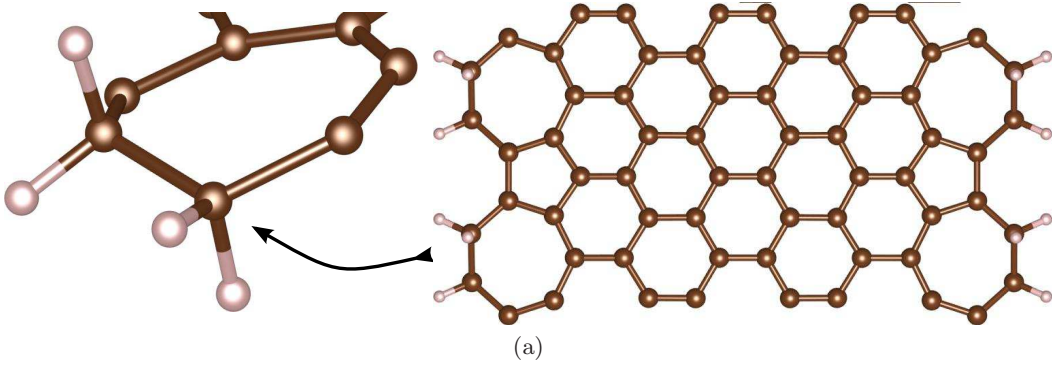}\\
(a)\\
\includegraphics[scale=0.32]{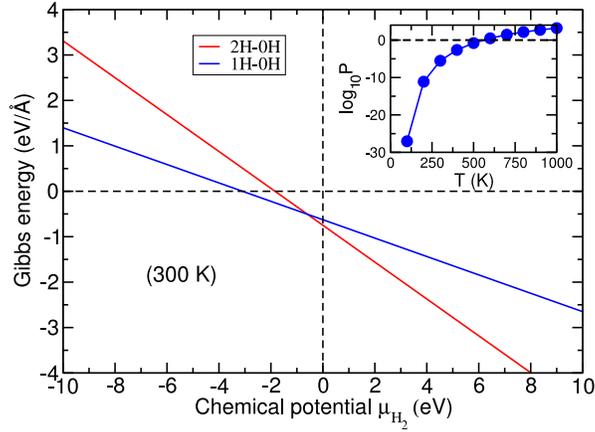}\\
(b)
\end{center}

\caption{\label{fig1} (Color online) (a) Reconstructed edge GNR with 2H termination. Brown (dark in print) balls are C atoms and white (light in print) balls indicate H atoms. The close-up of the edge structure is also shown; (b) Gibbs free energy calculated for 12 rows-reczag. Both 1H and 2H terminations with respect to bare reczag are presented. In the inset, transition pressures as a function of temperatures are shown. $P^{0}$ is the reference pressure taken to be 0.1 bar.}

\end{figure*}

The dangling bonds of edge carbon atoms are extremely reactive and need to be saturated. Among the several possible ways of edge termination, we concentrate on hydrogen termination at the edges as this seems to be one of the most stable configurations due to its planar structure. We have considered the reczag edge termination by one (1H) and two (2H) hydrogen atoms attached to each edge carbon atom for nanoribbons of width 4 to 12 rows. Figure \ref{fig1} shows the optimized geometry of a reczag edge with 2H termination. 

The C atoms in the middle of the ribbons rearrange themselves after geometry optimization and the bond lengths come out to be very close to the C-C bond distances of bulk {\g}
(1.42 {\AA}). However, the creation of edge reconstruction and H termination significantly
changes the C-C bond distances near the edge. In a bare reczag edge without H termination, the
edge carbon atoms form a triple bond with a bond length of 1.25 {\AA}. When this edge
is terminated by 1H, this bond length increases to 1.43 {\AA}. Both bare reczag edge and the 1H
terminated one have planar structures. In the case of 2H termination, the $sp^2$ planar structure becomes buckled, making an $sp^3$ like structure
with an angle 102$^{\circ}$ between the H and edge carbon atoms. The C-C bond length
increases further to 1.58 {\AA}. However, the C-C bond length of 1.42 {\AA} in $sp^2$ bonded graphene
is recovered in the middle of the ribbon.The edge carbon atoms are displaced from the plane
of the ribbon with one C atom shifting upwards and one downwards. H atoms
attached to these two carbon atoms also change their orientation to give rise to a twisted geometry as shown in
figure \ref{fig1}. To investigate the reason of twisting we have
done $\Gamma$ point (long wavelength) phonon calculations of 2H terminated reczag edge using frozen phonon method \cite{Martin}. We find that the structure
with 2 H atoms vertically placed on one another and connected to the edge C atom is not
stable, showing  two unstable modes involving the displacement of H atoms away from the
vertical position. For 2H termination, we therefore relaxed the structure again with these two
unstable modes frozen. The relaxed structure was stable and a twisted geometry (shown in
figure \ref{fig1}) was obtained due to the freezing of the above mentioned modes. It is obvious that the edge $sp^3$ structure
has substantial effects on the whole part of ribbons with smaller widths whereas the
structural distortion is not prominent in the middle of the wider ribbons.

For the termination of edges with 2 H atoms per C, the formation energies have been
defined as $ E_f = E(G2H) - [E(G1H) + n*E(H_2)] $, where $E(G2H)$ and $E(G1H)$ are the total energies for reconstructed graphene nanoribbons' edges terminated with 1H and 2H atoms per edge, respectively. $E(H_2)$ is the calculated energy for a $H_2$ molecule in the gas phase and $n$ is the number of $H_2$ molecules used to compensate the uneven number of H atoms. 

The calculated formation energies indicate that 2H terminated edge is probable to form. Also it is observed (data not shown) that 
the formation energies saturate after a width of eight rows as reported before \cite{bhandary} for a unreconstructed edge zigzag nanoribbon. Moreover, our calculation shows that the termination with H atoms will be spontaneous at T=0K. We have done test calculations by freezing the edge carbon bond lengths. The edge hydrogenation seems to be difficult if the C-C bond lengths are not allowed to relax.  Once the full geometry optimization is allowed, edge hydrogenation becomes probable. For the reconstructed edges, we find that a spontaneous formation of 2H terminated edges is possible rather than the 1H terminated ones for all widths.  This is in sharp contrast with the hydrogenation
at the unreconstructed ZGNRs studied earlier \cite{bhandary}.

In order to investigate the influence of finite
temperature/elevated gas pressure, we have calculated Gibbs free energies as a function of
the chemical potential of the hydrogen molecule, according to the following formula given
by Wassmann {\it et al.}\cite{wassmann}.

\begin{eqnarray*}
G_{H_{2}}=\frac{1}{2L}[E_{H_{2}}-(\frac{N_{H}}{2})\mu_{H_{2}}] \\
G_{H_{1}}=\frac{1}{2L}[E_{H}-(\frac{N_{H}}{2})\mu_{H_{2}}]  \\
E_{H_{2}} = E(G2H)-[E(G0H)+4E(H_{2})] \\
E_{H} = E(G1H)-[E(G0H)+2E(H_{2})] \\
\mu_{H_{2}} = H^{0}(T) - H^{0}(0) -TS^{0}(T) + k_{B}T~ln(\frac{P}{P^{0}})
\end{eqnarray*}

In the above equations, $\mu_{H_{2}}$, $H$, $S$, $P$ and $k_{B}$ are the chemical
potential, enthalpy, entropy, pressure and Boltzmann constant respectively and $N_{H}$ is
the number of H atoms attached at the edge.  The values for the entropies and enthalpies
are taken from the tabular data presented in Ref.~\cite{chase}.  $P^{0}$ is the
reference pressure taken to be 0.1 bar according to the tabular data. $E(G2H)$, $E(G1H)$ ,
$E(G0H)$ and $E(H_{2})$ are total energies for 2H, 1H and bare reconstructed nanoribbons
and hydrogen molecule respectively. The results are shown in figure~\ref{fig1} for
300 K.  The Gibbs free energy is normalized by 2L, where L is unit cell length.  The
stability of 2H and 1H terminated reconstructed nanoribbons  is shown with respect to bare
reconstructed nanoribbons . The zero temperature calculation shows that they are always favored
compared to the bare nanoribbons. Also, 2H terminated ribbons are more stable than the 1H ones. With the inclusion of temperature and pressure effects, it is observed that at low pressure, 1H
terminated edge can be stabilized over 2H terminated edge but after a certain pressure, 2H
edge becomes more stable. We call this cross over point as the transition point. The pressure required to reach this transition point increases with temperature as shown in the inset of figure~\ref{fig1}. The values of transition point pressures also suggest the possibility of the formation of a 2H terminated reczag edge at room temperature and ambient pressure. 
\begin{figure}[h]
\begin{center}
\includegraphics[scale=0.32]{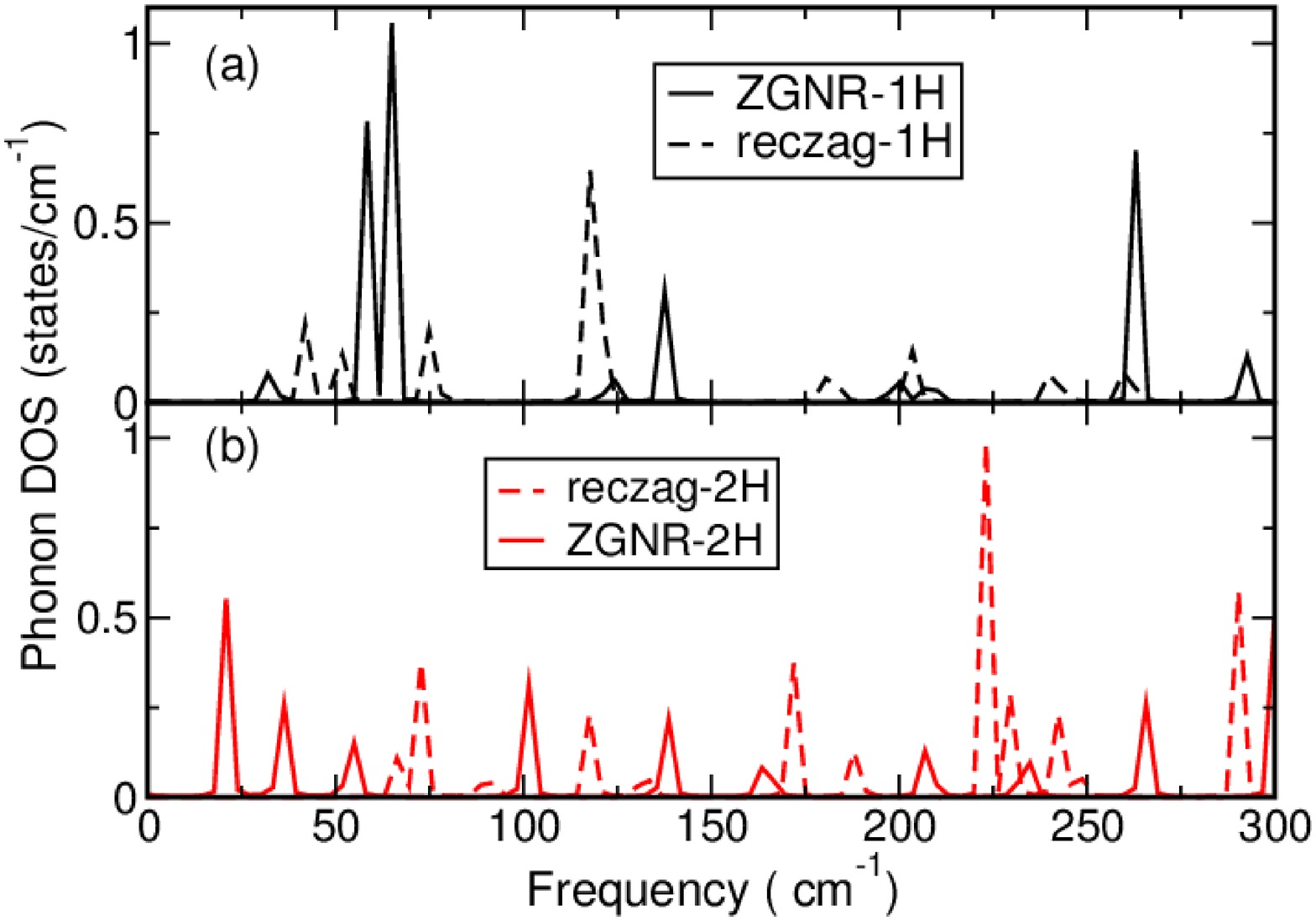}

\caption{(Color online) Zone center phonon DOSs plotted as a function of frequency for (a) 1H terminated ZGNR and reczag structures and (b) 2H terminated ZGNR and reczag structures. }

\label{fig:phonon}
\end{center}
\end{figure}

The calculated electronic structures (data not shown) for 1H and 2H terminations show the
presence of finite density of states at Fermi energy originating mostly from the $p_z$ orbitals of the C atoms next to the edge C atoms. However, the magnetic moment is lost due to the saturation of C-C bonds at the edges,  as seen earlier \cite{koskinenprl,koskinen2,wassmann}. Unreconstructed ZGNR with 1H and 2H terminated edges have finite magnetic moments and metallicity~\cite{son, bhandary}. In contrast to those, both 1H and 2H terminated reczag edges are non magnetic in the present case.

Now, we show the comparison of the phonon densities of states of 1H and 2H terminated unreconstructed and
reconstructed edge structures. The results are shown in figure \ref{fig:phonon} where only
optical phonons are displayed. From the figure it is clear that due to edge
reconstruction, the lowest optical phonon modes are hardened. This hardening is however
enhanced in the case when the edge is terminated with 2H compared to that of 1H.
 
%It should be noted that $\Gamma$ point phonons of ZGNR with N dimers are generally written
%as $$\Gamma = N(A_g\oplus B_{1g}\oplus B_{1u}\oplus B_{2g}\oplus B_{2u}\oplus B_{3u}),$$
%where the six modes inside the bracket are equivalent to six $\Gamma$-point modes in
%graphene \cite{phmodes}. These modes constitute six fundamental modes, and for a ZGNR with
%N dimers, there are 6N-6 overtones( each fundamental mode has N-1 overtones). The above
%irreducible representations consider D$_{2h}$ symmetry of a ideal ZGNR. However because of
%structural relaxation, we observe a lower symmetry and therefore it becomes difficult to
%identify six fundamental modes. Termination with hydrogen makes the situation even more
%complicated. A detailed discussion on the fundamental modes and overtones will follow in a future communication.

\subsection{Edge termination with Fe} 
\begin{figure}[h]
\begin{center}
\includegraphics[scale=0.28]{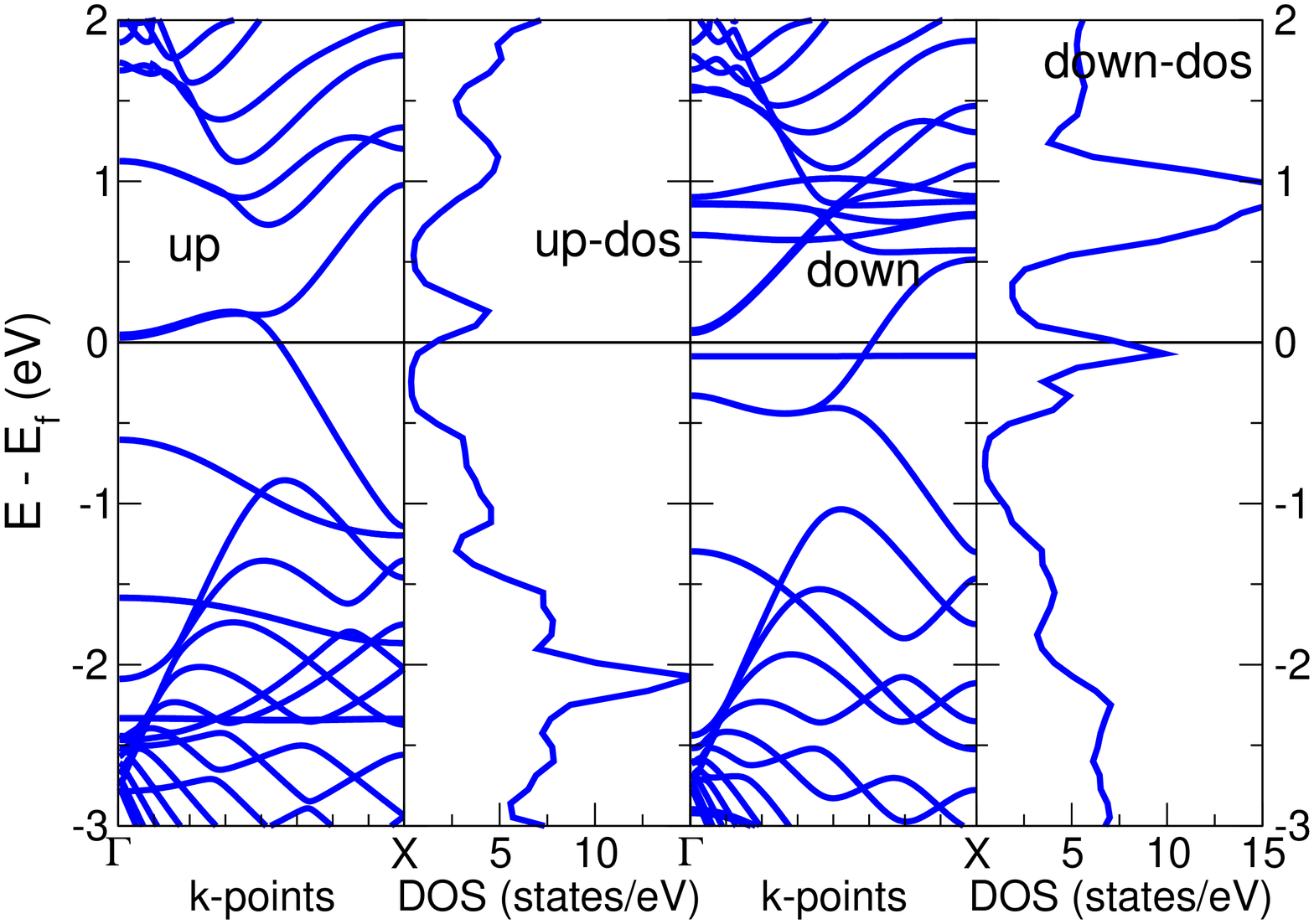}
\includegraphics[scale=0.28]{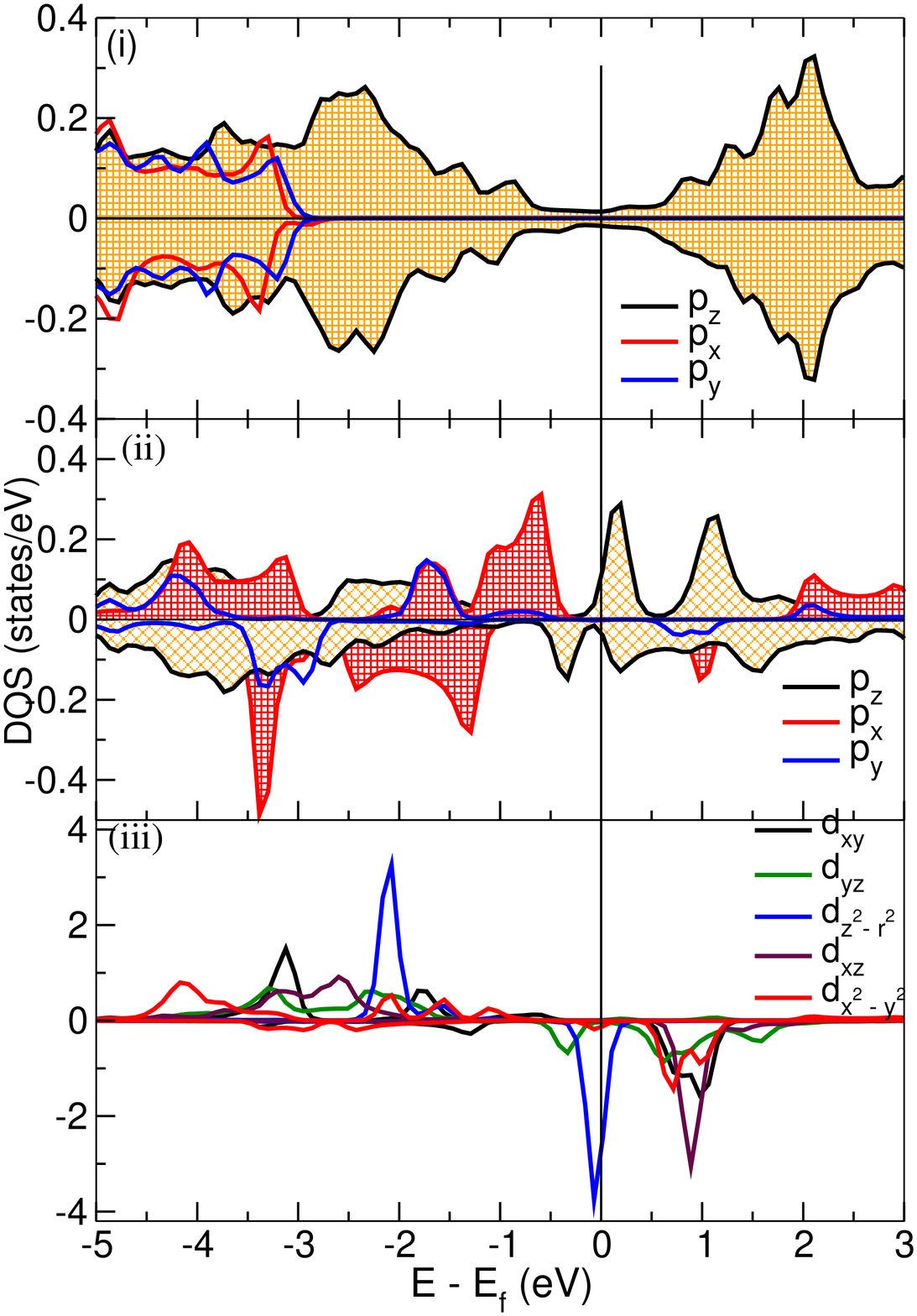}

\caption{(Color online) (left) Total DOSs and band structures for an Fe decorated 12 rows reczag edge. Both
spin-up and spin-down states are
shown. (Right) Spin-polarized site and orbital projected DOSs for (a) C atoms in the middle of the ribbon, (b) edge C atoms and (c) Fe atom. } \label{band-dos}

\end{center}
\end{figure}
The edge reconstruction destroys the magnetism of C atoms at the edges as the flat bands near or on
Fermi level for unreconstructed GNRs are now highly dispersive due to increased
hybridization. To introduce magnetism and also to observe magnetic interaction through
graphene lattice, we decorated the edges with Fe atoms. Our
geometry optimization shows that the Fe atom placed in between two heptagons is favorable
over top-hexagon or top-pentagon positions by at least 27 meV/C atom.  The calculated
formation energies of Fe terminated reczag edge are around 2.6 eV when the chemical potential of
Fe is taken as the same as bulk bcc Fe. 
The formation energy $E_{f}$ of a metal atom at the edge, is defined as
$E_{f} = E(Met_N+reczag) - [N*E(Met) + E(reczag)]$,
where $E(Met_N+reczag)$ is the total energy of geometry optimized Metal+reconstructed edge graphene nanoribbon, $E(reczag)$ is the total energy for the optimized geometry of reconstructed edge GNR and $E(Met)$ is
the chemical potential of the metal calculated in its bulk phase.  $N$ is the number of
metal atoms in the unit cell. It should be noted that a recent paper
\cite{haiping} reports a negative formation energy (indication of spontaneous formation) for Fe decoration at the zigzag or armchair edges when the chemical potential is taken from Fe in the atomic phase. So, the formation of Fe decorated edges highly depends on the Fe reference level.
\begin{figure}[h]
\begin{center}
\includegraphics[scale=0.3]{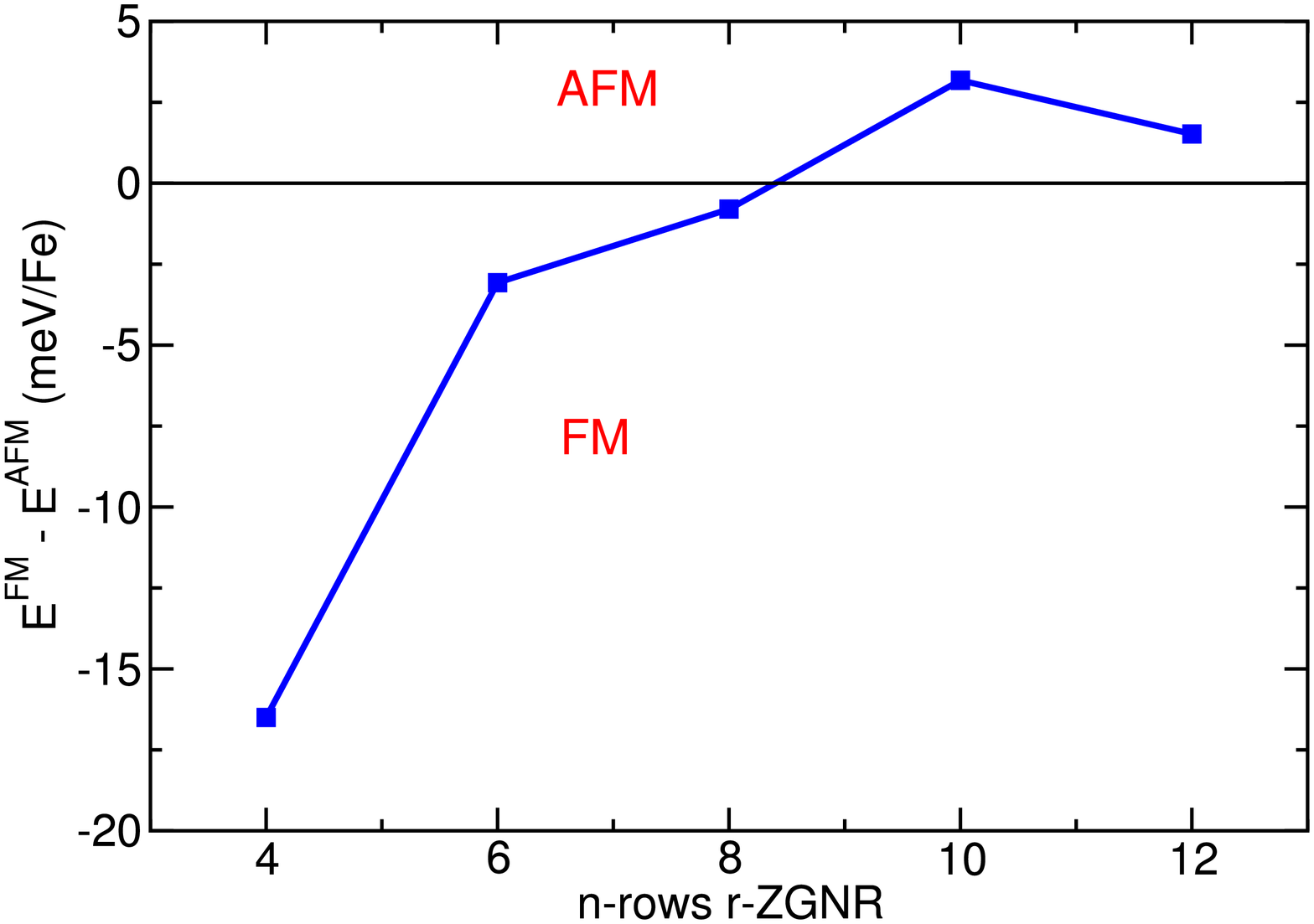}
\includegraphics[scale=0.2]{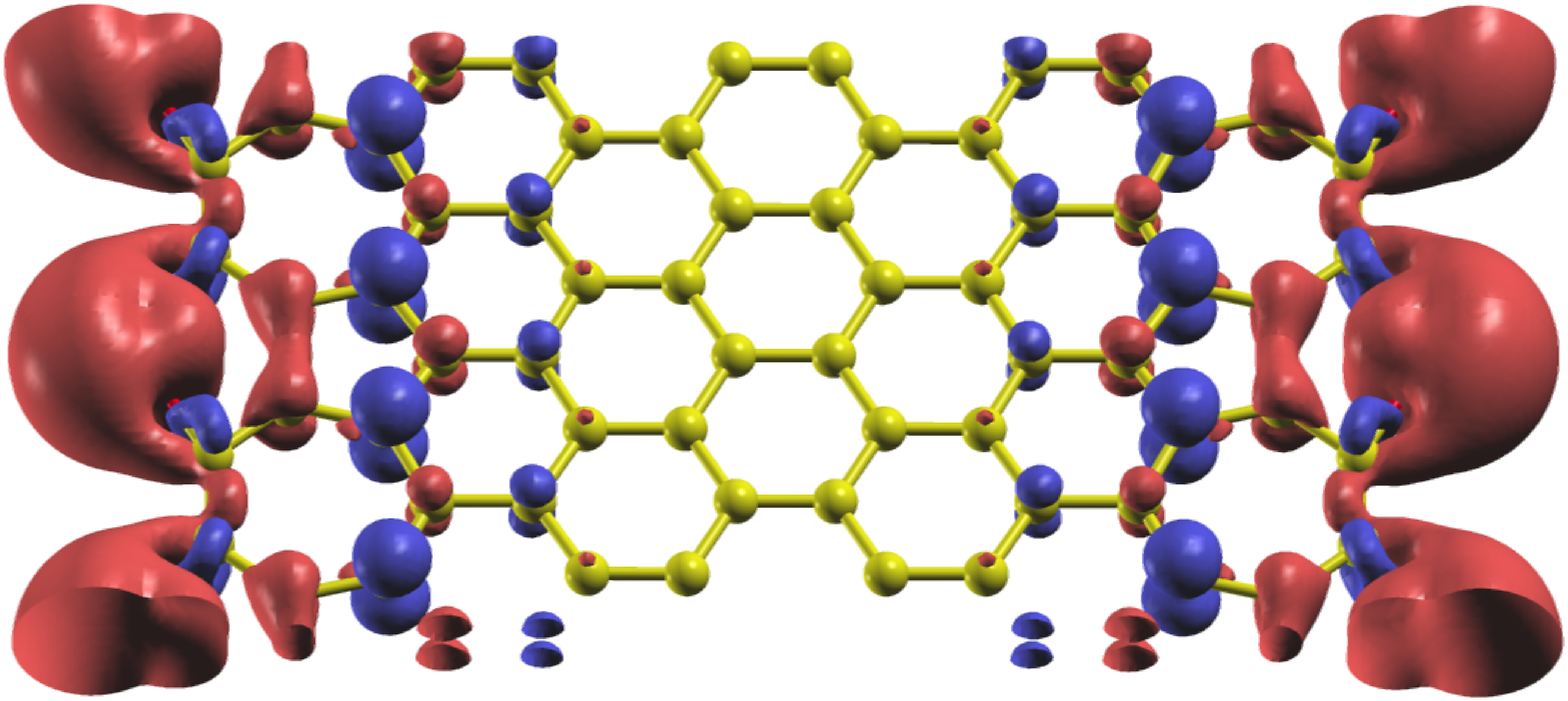}

\caption{(Color online) (top) Inter-edge exchange coupling (total energy difference between FM and AFM coupling across the edges) plotted as a function of ribbon width, (bottom) Spin density isosurfaces of 8-rows Fe doped reczag edge for a FM coupling across the edges. Red and blue colors represent spin-up and spin-down densities respectively.} \label{femag}

\end{center}
\end{figure}

Figure \ref{band-dos} shows the band structure and DOSs for an Fe edge-decorated 12 rows reczag edge. The spin down channel shows a localized peak in the total DOS (non dispersive band in the band structure plot) just below the Fermi level, contributed by Fe. The orbital projected DOSs on Fe (shown in panel (c) at the right of Figure \ref{band-dos}) clearly indicate that this localized peak has a  $d_{z^2-r^2}$ character in the spin-down channel. The spin-up channel of Fe is completely filled and has a negligible contribution within 2 eV below the Fermi level. The dispersive states in this energy range in the band structure arise mainly from edge C atoms. Projected DOSs on C atoms at the edge (panel (b)) are quite different from the ones in the middle (panel (a)) of the nanoribbon. The dominance of $p_z$ character within a considerable energy range around Fermi level in panel (a) is similar to what is observed for bulk graphene. The edge C atoms are spin-polarized and a strong contribution of in-plane $p_x$ 
and $p_y$ orbitals are seen in the similar energy range. 

If we look at the structural changes owing to the addition of Fe atoms, it has effect on
heptagons only as we have seen for the hydrogenated reczag edges. But the effect of Fe on the
graphene lattice is longer ranged as evident from the spin density isosurfaces shown in
figure \ref{femag}. The local moment on Fe is around 3.5 $\mu_{B}$ for all widths
considered. The Fe moments are quite robust and they are independent of width as well as
on the exchange coupling (FM or AFM) across the two edges of the nanoribbons. The closest
C atoms, to which Fe is bonded, have induced moments antiparallel to the Fe ones and same
for their same sublattice carbons, whereas other sublattice carbon atoms have in phase
magnetization.  As we travel towards the center of the ribbon, this effect is decreased.
The opposite magnetization in different sublattices is clearly visible throughout the
ribbon for the AFM coupled edges whereas for FM coupling, the magnetization disappears in
the middle of the nanoribbon due to the cancellation of induced moments.

We have done a calculation to check the possibility of monohydrogenation of a 12 rows reczag GNR already functionalized with Fe. We have found that it's comparatively less favorable (formation energy of -0.29 eV compared to -0.63 eV in the absence of Fe) to have a 1H termination in the presence of Fe. This is expected as Fe is already bonded at the edge and the H atom does not have enough room for bonding as it was possible in the absence of Fe.

\begin{figure}[h]
\begin{center}
\includegraphics[scale=0.15]{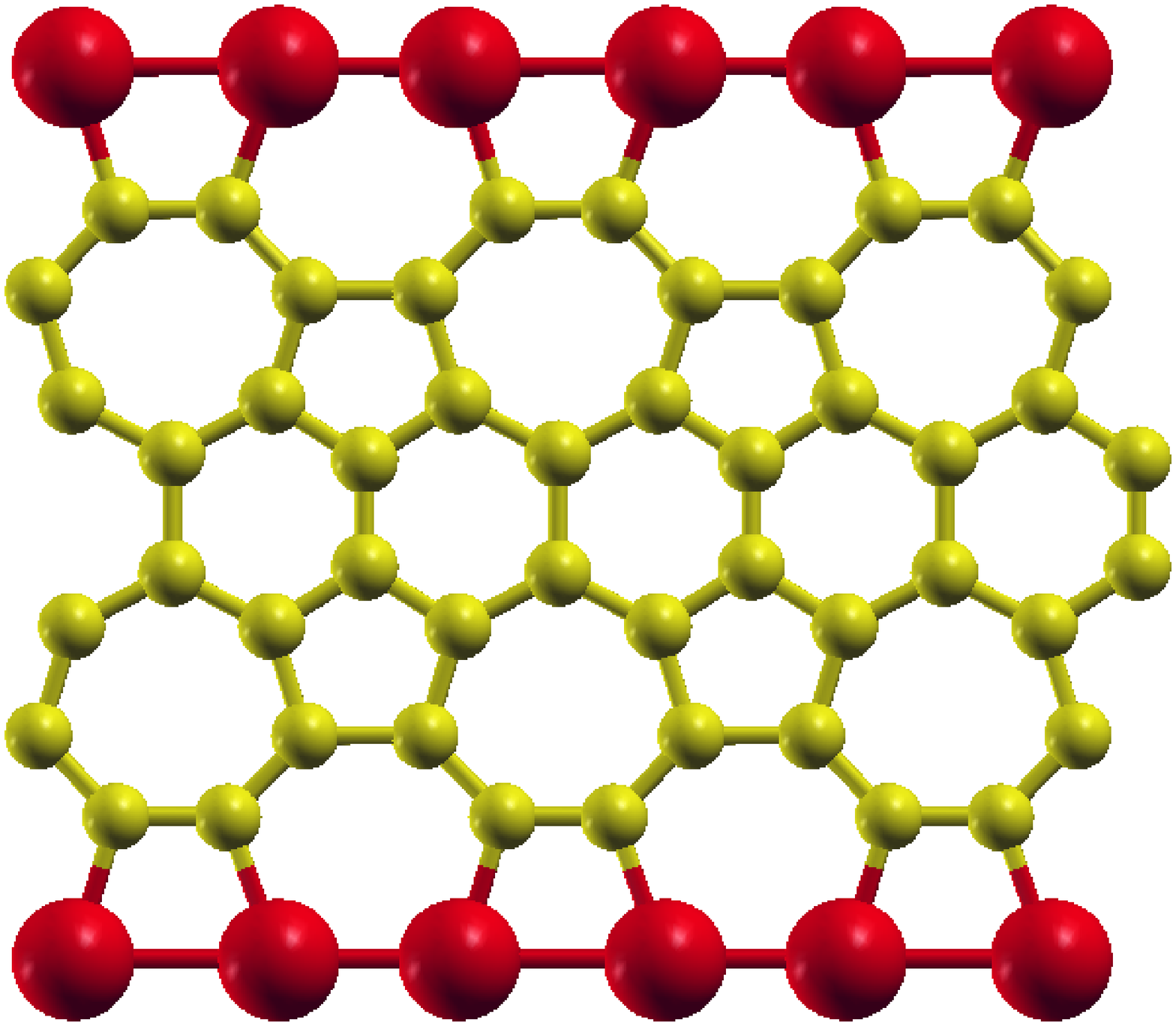}
\includegraphics[scale=0.15]{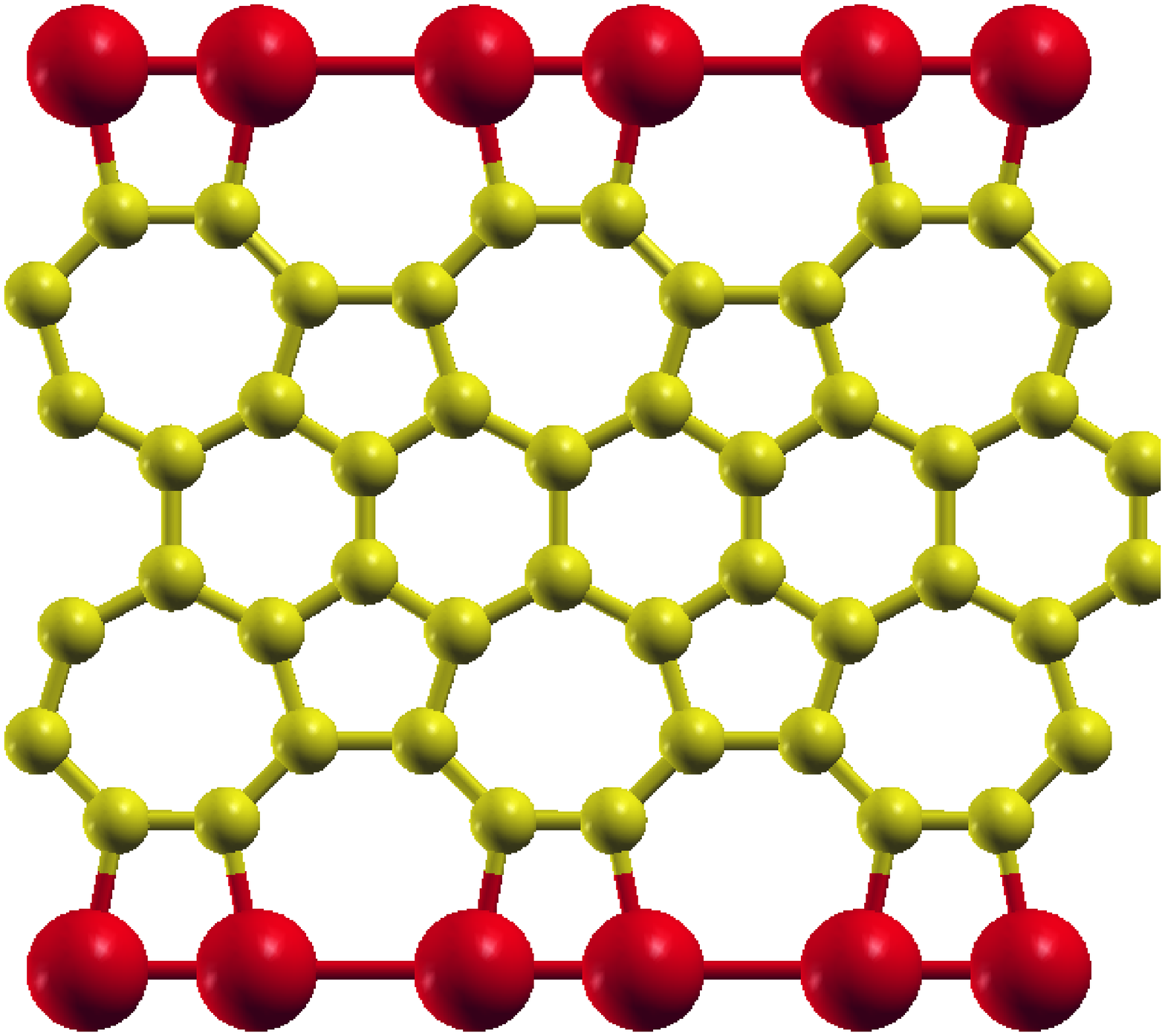}

\caption{(Color online) The structures of two dimerized chains indicated as Dimer 1 (left)
and Dimer 2 (right) in Table~\ref{tab4}. In the figure, red and yellow balls indicate Fe
and C atoms respectively.} \label{fig:dimers}

\end{center}
\end{figure}

The exchange coupling (measured by the total energy difference between FM and AFM coupling
across the edges) is plotted against the width of the nanoribbon and is shown in Figure
\ref{femag}.  For 4 rows, there is a relatively strong FM coupling. With the increase in
width, this coupling decreases and changes to an AFM one after 8 rows. The AFM coupling
tends to decrease afterwards. %In short, there is a tendency of an oscillatory exchange
%coupling across the edges. 
Interesting features of interedge coupling has been reported recently
\cite{haiping} for Fe and Co decorated armchair nanoribbons, where an oscillatory exchange coupling has been observed. The analogy with interlayer
exchange coupling in case of magnetic multilayers coupled through nonmagnetic layers \cite{iec} was made in that paper. The period of the coupling was analyzed in connection to Fermi
surface nesting vectors.
 
%**********************************************
Finally, we discuss the possibility of the formation of a dimerized Fe chain along the
edges of the nanoribbons as this issue has been explored in case of unreconstructed 6-6 nanoribbons \cite{haiping}. 
This situation is different from the above discussions where the density of edge Fe atoms was considered to be lower. Here we allow the Fe atoms to bind to the edge C atoms of the heptagon in contrast to the previous case, where one Fe atom was allowed to sit in between the heptagons and hence, the Fe atoms were far enough to have the possibility of dimerization.
To study the dimerization, we started the geometry optimization from two
different configurations of Fe dimers and have obtained the ground state geometries as
shown in figure \ref{fig:dimers}. One can observe two types of dimer structures, one formed
between the Fe atoms connected to the carbon atoms of a heptagon (second structure of
figure \ref{fig:dimers}) and the second type is formed between the Fe atoms connected to
carbon atoms belonging to adjacent heptagons. Both the dimers are stable with respect to
non-dimerized Fe termination. Table~\ref{tab4} shows the energetics, structure and
magnetism for the two cases. It is evident that Fe dimers connected to the same heptagon
is favorable over the other one by an energy of 0.06 eV. The bond length of this dimer
comes out to be 2.14~\AA~whereas the other structure yields a bond length of 2.3~\AA. The
local magnetic moments on Fe are same for these two structures and are close to 3 $\mu_{B}$. It is interesting to note that the moments are reduced compared to the non-dimerized Fe chains, where the hybridization between the Fe-d orbitals was weaker due to larger separations. The electronic structure and corresponding magnetic exchange coupling for these dimerized structures will be discussed in a future communication.

\begin{table}
\begin{tabular}{|c|c|c|c|c|}
\hline
Dimer& $\Delta E$ (eV) & Fe-Fe & Fe-Fe & moment \\
 & & distance 1 (\AA) & distance 2 (\AA) & ($\mu_{B}$)\\
\hline
\hline
1&-0.23 &2.61&2.30&2.94\\
\hline
2&-0.29 &2.77&2.14&2.94\\
\hline
\end{tabular}

\caption{Formation energies ($\Delta E$) of two types of Fe dimers in the chain with
respective to non-dimerized structure. The corresponding bond lengths (short and long) of
the Fe dimers are shown for the two structures shown in figure \ref{fig:dimers} along with
the local magnetic moments at Fe sites.}

\label{tab4}
\end{table}

%**************************************************************************
\section{Conclusions}

In this paper, chemical functionalization at the edges of reconstructed zigzag graphene nanoribbons has been studied by ab initio
density functional theory. Reconstructed edges do not show any magnetism and have a metallic behavior.  From our calculations, it is seen that both single and
dihydrogenated reczag edges are probable to form, as observed in previous theoretical calculations.. Unlike unreconstructed ZGNRs, dihydrogenation
is always favorable over monohydrogenated reczag edges, independent of the width of the
nanoribbons under ambient conditions. We have also shown that at finite temperatures,
the hydrogen pressure dictates the formation of mono- or di- hydrogenated edges.  Our
phonon calculations reveal a peculiar geometry at the dihydrogenated edges. Moreover, it
has been found that the lowest optical phonon modes are hardened due to edge
reconstruction. To render magnetism in
reczag edges, we have decorated the edges by Fe chains. The interedge magnetic coupling varies between ferromagnetic and antiferromagnetic ones with the variation of the width of the nanoribbons
with robust localized moments residing at the Fe sites. Finally, we show that the Fe atoms
in the chain along an edge prefer to be in a dimerized configuration.  

\section*{Acknowledgements}

SH would like to acknowledge Indo-Swiss grant for financial support (No: INT/SWISS/P-
17/2009).  BS acknowledges Swedish Research Links programme under VR/SIDA, G\"{o}ran
Gustafssons Stiftelse, Carl Tryggers Stiftelse and KOF initiative by Uppsala University
for financial support. We thank SNIC-UPPMAX, SNIC-NSC and SNIC-HPC2N computing centers under Swedish
National Infrastructure for Computing (SNIC) for granting computer time. O.E. acknowledges
support from ERC and the KAW foundation.

%\section*{References}
\bibliographystyle{elsarticle-num}

 \end{document}